\documentclass[12pt]{article}
\usepackage{amsmath}
\usepackage{amssymb}
\usepackage[dvips]{graphicx}
\usepackage{cite}

\numberwithin{equation}{section}

\begin{document}
\baselineskip=18pt
\begin{titlepage}
\begin{flushright}
{\small KYUSHU-HET-116}
\end{flushright}
\begin{center}
\vspace*{11mm}

{\Large\bf%
Probing neutrino masses and tri-bimaximality\\[2mm]%
with lepton flavor violation searches%
}\vspace*{8mm}

Kentaro Kojima$^{a,}$\footnote{E-mail: kojima@rche.kyushu-u.ac.jp} 
and 
Hideyuki Sawanaka$^{b,}$\footnote{E-mail: sawanaka@higgs.phys.kyushu-u.ac.jp}
\vspace*{5mm}

{\it $^a$ Center for Research and Advancement in Higher Education,\\
Kyushu University, Fukuoka 810-8560, Japan\\
$^b$ Department of Physics, Kyushu University, Fukuoka 812-8581, Japan}

\vspace*{3mm}

{\small (January, 2009)}
\end{center}
\vspace*{5mm}

\begin{abstract}\noindent%
We examine relation between neutrino oscillation parameters and prediction 
of lepton flavor violation, in light of deviations from tri-bimaximal 
mixing. Our study shows that upcoming experimental searches for lepton 
flavor violation process can provide useful implications for neutrino mass 
spectrum and mixing angles. With simple structure of heavy right-handed 
neutrino and supersymmetry breaking sectors, the discovery of 
$\tau\to \mu\gamma$ decay determines neutrino mass hierarchy if large 
(order 0.1) reactor angle is established. 
\end{abstract}
\end{titlepage}
\newpage

\section{Introduction}
\label{sec:int}

In the last decades, one of the most striking developments in particle 
physics beyond the standard model (SM) is the experimental 
establishment~\cite{SK98} of neutrino masses and the large mixing property, 
which is quite different from the small mixing in the quark sector. 
Neutrino oscillation experiments have revealed neutrino mass-squared 
differences and its flavor mixing angles. Notably, a simple form of mixing 
matrix, referred to as tri-bimaximal mixing, is well descriptive of the 
observed mixing structure~\cite{tbm}. Vast numbers of flavor models have been 
proposed in order to derive the tri-bimaximal mixing~\cite{model}; thus, from 
a theoretical viewpoint, it is one of the most important subjects to realize 
difference between tri-bimaximal and observed mixing angles. 
In addition, recent results of the 
three-flavor global data analysis~\cite{Gfit,13mix} indicate non-zero 
$\theta_{13}$; as the best-fit value, not so small one 
$\sin\theta_{13}\simeq {\cal O}(0.1)$ is 
obtained. Therefore, it seems interesting to examine deviations of neutrino 
mixing angles from the tri-bimaximal pattern~\cite{deviation}, which leads to  
$\sin^2\theta_{12}=1/3$, $\sin^2\theta_{23}=1/2$, and $\sin^2\theta_{13}=0$, 
for the coming future precise experiments. 

The present knowledge of neutrino parameters (i.e. masses, mixing angles 
and phases) is not only important for low-energy characters of neutrinos, 
but also a key ingredient of the origin of flavor structure in the SM 
fermions. It is thus meaningful to make clear the relation between 
neutrino parameters and high-energy phenomenologies. Among them, charged 
lepton flavor violation (LFV) process would give an intriguing 
clue, since supersymmetry (SUSY) and the seesaw mechanism~\cite{seesaw} could 
enhance LFV as reachable in near future experiments~\cite{lfv}. The seesaw 
mechanism naturally provides desired neutrino mass scale, and predicted 
LFV fractions are affected by low-energy neutrino parameters and 
heavy Majorana masses via SUSY breaking terms~\cite{rdep,lfvfl}. 

In this Letter, we examine the relation between neutrino parameters and 
LFV prediction, in light of the tri-bimaximal mixing and the recent precision 
oscillation data. We use a particular parametrization~\cite{tmin} of 
the MNS matrix~\cite{mns}, 
where the tri-bimaximal mixing is taken as its zeroth order approximation, 
and give a detailed analysis of LFV prediction especially for a simple case of 
right-handed Majorana mass and SUSY breaking structures. 
Our study shows that upcoming experimental LFV searches 
provide useful implications for neutrino mass spectrum and mixing angles. 
If we possess simple structure of heavy Majorana masses, future discovery of 
$\tau\to \mu\gamma$ implies that inverted hierarchy (IH) and 
quasi-degenerate (QD) neutrino mass spectra are inconsistent with 
a large reactor angle of order 0.1, though normal hierarchy (NH) is still 
allowed. 

The Letter is organized as follows. In Section~\ref{sec:Neu}, we introduce 
a useful parametrization~\cite{tmin} of the MNS matrix for examining 
how the mixing angles deviate from the tri-bimaximal ones. In 
Sections~\ref{sec:LFV} and~\ref{sec:LFV-TB}, relation between neutrino 
parameters and LFV prediction in a literature of the minimal 
supersymmetric standard model (MSSM) with the seesaw mechanism is studied. 
Detailed analysis is given in Section~\ref{sec:LFV-TB} by using the 
parametrization. Section~\ref{sec:clarif} is devoted to study implications 
for neutrino parameters with future LFV searches. We summarize our result 
in Section~\ref{sec:conclusion}. 

\section{Neutrino parameters and tri-bimaximal mixing}
\label{sec:Neu}

Current data obtained by neutrino oscillation experiments is consistent 
with a simple mixing structure called tri-bimaximal mixing~\cite{tbm}. 
We adopt a particular parametrization proposed in Ref.~\cite{tmin} 
to describe the MNS matrix by deviations from exact tri-bimaximal mixing angles. 
It is helpful to systematically analyze the neutrino tri-bimaximality. 

In the basis where the charged lepton mass matrix is diagonal, 
neutrino mass matrix is given by $M_\nu= U {\cal D}_m U^T$, where 
${\cal D}_m=diag(m_1, m_2, m_3)$ with positive neutrino mass eigenvalues 
$m_{1, 2, 3}$ and $U$ is the unitary lepton mixing matrix. Three mixing 
angles and phases are involved in the matrix $U$; 
using the standard parametrization~\cite{PDG08}, it can be expressed as 
\begin{eqnarray}
U &=&
  \begin{pmatrix}
    c_{12}c_{13}&s_{12}c_{13}&s_{13} e^{-i \delta}\\
    -s_{12}c_{23}-c_{12}s_{23}s_{13} e^{i\delta}&
    c_{12}c_{23}-s_{12}s_{23}s_{13}e^{i\delta}&
    s_{23}c_{13}\\
    s_{12}s_{23}-c_{12}c_{23}s_{13}e^{i\delta}&
    -c_{12}s_{23}-s_{12}c_{23}s_{13}e^{i\delta}&
    c_{23}c_{13}
  \end{pmatrix}
P_M, 
\label{mnspdg}
\end{eqnarray}
where 
$c_{ij}\equiv\cos\theta_{ij}$, $s_{ij}\equiv\sin\theta_{ij}$, 
and $\delta$ is the Dirac CP violating phase. $P_M$ stands for the diagonal 
phase matrix which involves two Majorana phases. 

Recent progress in neutrino experiments greatly increase data for 
neutrino masses and mixing angles. The updated result of the three-flavor 
global data analysis~\cite{Gfit} indicates the following best-fit values with 
3$\sigma$ intervals of three (solar, atmospheric, reactor) mixing angles 
and two (solar, atmospheric) mass squared differences: 
\begin{eqnarray}\label{mixex}
 \sin^2\theta_{12}\;=\;0.304^{+0.066}_{-0.054}, &&
 \sin^2\theta_{23}\;=\;0.50^{+0.17}_{-0.14},\qquad 
 \sin^2\theta_{13}\;=\;0.010\ (\le 0.056),\\[2mm]\notag
\Delta m_{\rm sol}^2&\equiv& \, m_2^2-m_1^2\ \;=\;
(7.65\pm^{0.69}_{0.60})\times 10^{-5}~{\rm eV}^2,\\\label{msqdiff}
\Delta m_{\rm atm}^2&\equiv&|m_3^2-m_1^2|\;=\;
(2.40\pm^{0.35}_{0.33})\times 10^{-3}~{\rm eV}^2. 
\end{eqnarray}
The Dirac phase $\delta$ has not yet been constrained 
by the experimental data. 

The flavor structure could be determined by profound principles such as 
flavor symmetry in a high-energy regime. Although the origin of the flavor is 
unrevealed, the current neutrino mixing angles are known to be consistent 
with a simple mixing matrix $U_{\rm TB}$, 
\begin{eqnarray}
    U_{\rm TB} &=&
({\cal R}_{\rm TB})_{23}({\cal R}_{\rm TB})_{12}\;=\;
{1\over \sqrt{6}}
\begin{pmatrix}
2&\sqrt{2}&0\\
-1&\sqrt{2}&\sqrt{3}\\
1&-\sqrt{2}&\sqrt{3}
\end{pmatrix}, 
\label{tbm}
\end{eqnarray}
where two (non-diagonal) rotational matrices are given by 
\begin{eqnarray}
  ({\cal R}_{\rm TB})_{23}&=&    
\begin{pmatrix}
  1&0&0\\
  0&{1/\sqrt{2}}&{1/\sqrt{2}}\\
  0&{-1/\sqrt{2}}&{1/\sqrt{2}}
\end{pmatrix}, \qquad 
({\cal R}_{\rm TB})_{12}\;=\;
\begin{pmatrix}
  {2/\sqrt{6}}&{1/ \sqrt{3}}&0\\
  {-1/ \sqrt{3}}&{2/ \sqrt{6}}&0\\
  0&0&1
\end{pmatrix}. 
\end{eqnarray}
Although the matrix $U_{\rm TB}$ approximately describes the MNS 
matrix~\eqref{mnspdg}, it is an interesting subject to investigate 
the difference between them in reality. For this purpose, the following 
unitary matrix is useful to parametrize the MNS mixing structure: 
\begin{eqnarray}
      U &=&
({\cal R}_{\rm TB})_{23}
  \begin{pmatrix}
     c_x c_z&s_x c_z&s_z e^{-i \delta}\\
     - s_x c_y - c_x s_y s_z e^{i\delta}&
     c_x c_y - s_x s_y s_z e^{i\delta}&s_y c_z\\
     s_x s_y - c_x c_y s_z e^{i\delta}&
     - c_x s_y - s_x c_y s_z e^{i\delta}&c_y c_z 
  \end{pmatrix}
({\cal R}_{\rm TB})_{12} 
P_M, 
\label{mnstb}
\end{eqnarray}
where $c_w$ and $s_w$ denote $\cos\epsilon_w$ and $\sin\epsilon_w$ 
($w=x, y, z$), respectively.%
\footnote{One can easily find that $\epsilon_z=\theta_{13}$ 
from~\eqref{mnspdg} and~\eqref{mnstb}, owing to $\theta_{13}=0$ in the 
tri-bimaximal limit.} 
In the limit where $\epsilon_{x,y,z}\to 0$, $U$ in 
\eqref{mnstb} goes back to $U_{\rm TB}$ except $P_M$. 
With this parametrization, experimental data \eqref{mixex} indicates 
\begin{eqnarray} 
-0.092\le \;\epsilon_{x}\;\le 0.038,\qquad 
-0.14\le \;\epsilon_{y}\;\le 0.17,\qquad 
|\epsilon_z|\; \le 0.239,
\label{devrange} 
\end{eqnarray}
in its $3\sigma$ ranges. Since the deviation parameters are much suppressed 
than ${\cal O}(1)$, expansions around $\epsilon_{x,y,z}=0$ could give 
a good approximation to physical quantities related to lepton mixing angles. 

\section{LFV in MSSM with type-I  seesaw}
\label{sec:LFV}

Let us study LFV prediction in MSSM with heavy right-handed Majorana 
neutrinos for the type-I seesaw mechanism~\cite{seesaw}. 
The relevant part of the MSSM superpotential is given by 
\begin{eqnarray}
  W_{\rm lepton}&=& 
L_i(Y_e)_{ij}\bar e_jH_d+L_i(Y_\nu)_{ij}\bar \nu_jH_u
+{1\over 2}\bar \nu_i(M_R)_{ij}\bar \nu_j, 
\end{eqnarray}
where $Y_e$ and $Y_\nu$ are charged lepton and neutrino Yukawa matrices. 
Superfields $L_i$, $\bar e_i$, $\bar \nu_i$ ($i=1,2,3$) and $H_{u(d)}$ include 
lepton doublets, charged leptons, right-handed neutrinos and up (down)-type 
Higgs doublet, respectively. $M_R$ gives Majorana mass matrix of the 
right-handed neutrinos, whose scale is assumed to be much larger than the 
electroweak scale ($\sim 10^{2}$~GeV). It is also noted that scale of $M_R$ 
is required to be smaller than the grand unified theory (GUT) scale 
in order to reproduce known solar and atmospheric neutrino mass scales 
unless $Y_\nu$ is much larger than ${\cal O}(1)$. 
Without loss of generality, we take a basis where $M_R$ is diagonal. 

SUSY breaking should be incorporated in realistic models, since it is not 
exact symmetry in Nature. We thus introduce SUSY breaking terms, which 
in general could be new sources of the flavor violation. Although several 
breaking scenarios have been proposed, one of the most economical and 
predictive ansatz is to assume the universal form of soft terms in a 
high-energy regime. In this case, universal SUSY breaking parameters are 
listed as scalar mass $m_0$, trilinear coupling $A_0$, gaugino mass $M_{1/2}$ 
and the Higgs bilinear coupling. We refer to these SUSY breaking 
parameters as their GUT scale values. 

The flavor violation in the supersymmetric sector is transmitted to SUSY 
breaking sector thorough renormalization group (RG) evolution between GUT and 
heavy Majorana mass scales. At a low-energy regime, one-loop diagrams with 
SUSY particles give leading corrections to the LFV processes; the branching 
fractions are approximately written as 
\begin{eqnarray}
  {\cal B}(\ell_j\to \ell_i+\gamma)&\simeq& {\alpha^3\over G_F^2m_S^8}
\left[{3m_0^2+A_0^2\over 8\pi^2v_H^2\sin^2\beta}\right]^2\tan^2\beta
|B_{ij}|^2,
\label{brap}
\end{eqnarray}
where $\alpha$ and $G_F$ are the fine-structure and the Fermi coupling 
constants, $v_H\simeq 174~{\rm GeV}$, and $\beta$ parametrizes the ratio 
between vacuum expectation values of the two Higgs scalars in $H_{u,d}$. 
The mass parameter $m_S$ is a typical mass scale of SUSY particles.  
Note that flavor indices only appear in $B_{ij}$ as 
\begin{eqnarray}
  B_{ij}&=&v_H^2\sin^2\beta\sum_{k=1}^3 (Y_\nu)_{ik}(Y_\nu^\dag)_{kj}
\ln{M_G\over {M_R}_k}, 
\label{fl}
\end{eqnarray}
where ${M_R}_k$ denotes the $k$-th eigenvalue of $M_R$, and 
$M_G\simeq 10^{16}$~GeV is the GUT scale. Thus the flavor 
dependence in LFV branching fractions is completely involved in~\eqref{fl}, 
that is SUSY breaking parameters do not affect the flavor structure in the 
approximate formula of LFV prediction~\eqref{brap}. 

Although it is generally difficult to reconstruct the combination of neutrino 
Yukawa matrix in~\eqref{fl} from low-energy data, if heavy Majorana neutrinos 
have an approximately degenerate mass $M_U$ and there are no large CP phases 
except the Dirac phase $\delta$ in the neutrino sector, the branching 
fractions~\eqref{brap} are tightly connected to neutrino parameters as 
\begin{eqnarray}
    {\cal B}(\ell_j\to \ell_i+\gamma)&\simeq& {\alpha^3\over G_F^2m_S^8}
\left[{3m_0^2+A_0^2\over 8\pi^2v_H^2\sin^2\beta}\right]^2\tan^2\beta
M_U^2\left(\ln{M_G\over M_U}\right)^2
|b_{ij}|^2, \\\label{cij}
&&  b_{ij}\;=\;\sum_{k=1}^3m_k(U^*)_{ik}(U^T)_{kj}. 
\end{eqnarray}
Note that $P_M$ in~\eqref{mnstb} disappears in the factor $b_{ij}$.
Here the low-energy neutrino parameters, namely their masses, mixing angles 
and the  Dirac phase, are completely involved in (\ref{cij}); thus the LFV 
branching fractions depend on SUSY parameters in a flavor independent manner.  
We will discuss more general cases with non-degenerate Majorana masses 
in the end of the next section.\footnote{One can also discuss relation between 
neutrino parameters and LFV prediction with other seesaw mechanisms than the 
conventional type-I~\cite{rode}. For instance, in the type-II seesaw 
scenario~\cite{seesawII}, neutrino masses in $b_{ij}$~\eqref{cij} are 
replaced by those squared, and mixing parameter dependence of LFV prediction 
can be studied as our analysis.}

\section{LFV prediction around tri-bimaximal mixing}
\label{sec:LFV-TB}

In this section, we discuss the relation between neutrino parameters 
and prediction of LFV processes. The parametrization~\eqref{mnstb} allows us to 
handle deviations from tri-bimaximal mixing in a systematic way. In a 
definite framework for high-energy theory, we analyze how LFV prediction 
depends on small deviation parameters with current neutrino oscillation data. 

\paragraph{Analytical results} 
Applying the parametrization~\eqref{mnstb} to 
describe the LFV prediction, $b_{ij}$ is explicitly written as 
\begin{eqnarray}\notag
b_{12}&=&{m_{12}\over 6\sqrt{2}}c_z(c_y-s_y)(2\sqrt{2}c_{2x}+s_{2x})
+{e^{i\delta}\over 6\sqrt{2}}c_zs_z(c_y+s_y)
\left[3m_{123}+m_{12}(c_{2x}-2\sqrt{2}s_{2x})\right],\\[2mm]
\notag
b_{13}&=&-{m_{12}\over 6\sqrt{2}}c_z(c_y+s_y)(2\sqrt{2}c_{2x}+s_{2x})
+{e^{i\delta}\over 6\sqrt{2}}c_zs_z(c_y-s_y)
\left[3m_{123}+m_{12}(c_{2x}-2\sqrt{2}s_{2x})\right],\\[2mm]\notag
b_{23}&=&{m_{123}\over 4}c_z^2(c_y^2-s_y^2)
-{m_{12}\over 12}(1+s_z^2)(c_y^2-s_y^2)
(c_{2x}-2\sqrt{2}s_{2x})
\\&&\hspace{3cm}
-{e^{i\delta}\over 12}m_{12}s_z\left[(c_y-s_y)^2
-(c_y+s_y)^2e^{-2i\delta}\right](2\sqrt{2}c_{2x}+s_{2x}), 
\label{cijfull}
\end{eqnarray}
where $c_{2x}=\cos2\epsilon_x$ and $s_{2x}=\sin2\epsilon_x$. Note that 
neutrino masses appear just as particular combinations  
\begin{eqnarray}
  m_{12}\;=\;m_2-m_1,\qquad m_{123}\;=\;2m_3-m_2-m_1. 
\end{eqnarray}
It is also noted that 
$b_{12}(\epsilon_y, \epsilon_z; \delta)
=-b_{13}(-\epsilon_y, -\epsilon_z; \delta)
=-b_{13}(-\epsilon_y, \epsilon_z; \delta\pm\pi)$ 
holds between $b_{12}$ and $b_{13}$; 
the relation is just the $\mu$-$\tau$ symmetric property in the 
tri-bimaximal limit. 

As mentioned in Section~\ref{sec:Neu}, the observed values of lepton mixing 
angles are consistent with the tri-bimaximal mixing pattern. Focusing on 
the LFV prediction with the tri-bimaximal limit  
$(\epsilon_x,\epsilon_y,\epsilon_z)\to (0,0,0)$ in~\eqref{cijfull}, we obtain 
\begin{eqnarray}\notag
{\cal B}(\mu\to e\gamma)&\propto&|b_{12}|^2\;\to\;
\Bigl({m_{12}\over 3}\Bigr)^2\;=\;{m_{12}^2\over 9},
\\\notag
{\cal B}(\tau\to e\gamma)&\propto&|b_{13}|^2\;\to\;
\Bigl({m_{12}\over 3}\Bigr)^2\;=\;{m_{12}^2\over 9},
\\\label{lfvtbl}
{\cal B}(\tau\to \mu\gamma)&\propto&|b_{23}|^2
\;\to\;
{1\over 16}\left(m_{123}-{m_{12}\over 3}\right)^2\;=\;
{m_{123}^2\over 16}-{m_{123}m_{12}\over 24}+{m_{12}^2\over 144}, 
\end{eqnarray}
where only $|b_{23}|^2$ depends on $m_{123}$ and $|b_{12}|^2=|b_{13}|^2$ 
holds. 

\begin{figure}[t]
\begin{center}
  \includegraphics[width=7cm,clip]{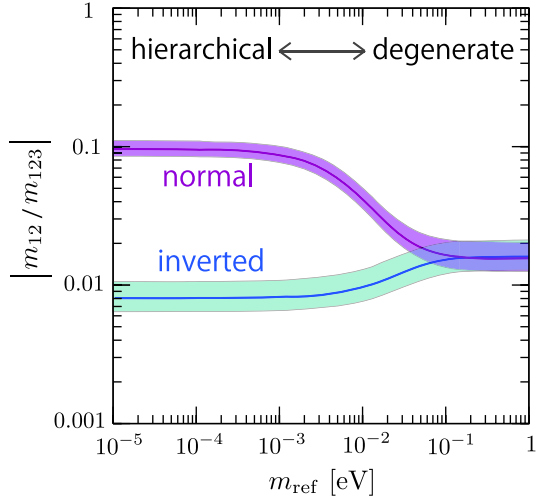}
\end{center}
\caption{Absolute values of the ratio between $m_{12}$ and $m_{123}$ are 
shown as functions of reference neutrino mass scale $m_{\rm ref}$; for the 
NH (IH) case we take $m_{\rm ref}=m_1(m_3)$. Colored bands indicate the 
predicted values with 3$\sigma$ ranges of input parameters 
$\Delta m_{\rm sol}^2$ and $\Delta m_{\rm atm}^2$. Lines in the bands 
correspond to the plot with the best-fit values in~\eqref{msqdiff}. 
  \bigskip}
\label{massdifs}
\end{figure}

Experimental values of solar and atmospheric neutrino mass differences 
indicate that $m_{123}$ is much larger than $m_{12}$. The ratio 
$\hat{m}\equiv m_{12}/m_{123}$ depends on the neutrino mass spectrum. 
For NH of neutrino masses ($m_1<m_2< m_3$), we obtain 
\begin{eqnarray}\label{mhatn}
  \hat{m}&=&
 {\sqrt{\Delta m_{\rm sol}^2+m_1^2}-m_1\over 
  2\sqrt{\Delta m_{\rm atm}^2+m_1^2}-\sqrt{\Delta m_{\rm sol}^2+m_1^2}-m_1} 
 \;\to \;{1\over 2}\sqrt{\Delta m_{\rm sol}^2\over\Delta m_{\rm atm}^2}, 
\end{eqnarray}
while for the IH case ($m_3< m_1<m_2$), we have 
\begin{eqnarray}\label{mhati}
  \hat{m}&=&
 {\sqrt{\Delta m_{\rm atm}^2+\Delta m_{\rm sol}^2+m_3^2}
 -\sqrt{\Delta m_{\rm atm}^2+m_3^2}\over 
  2m_3-\sqrt{\Delta m_{\rm atm}^2+\Delta m_{\rm sol}^2+m_3^2}-
  \sqrt{\Delta m_{\rm atm}^2+m_3^2}}
 \;\to \;-{1\over 4}{\Delta m_{\rm sol}^2\over\Delta m_{\rm atm}^2}.  
\end{eqnarray}
The last expressions in~\eqref{mhatn} and~\eqref{mhati} imply the massless 
limit of the lightest neutrino, where 
$\Delta m_{\rm sol}^2/\Delta m_{\rm atm}^2\simeq 0.04$ with the present data. 
We plot in Fig.~\ref{massdifs} the ratio $\hat{m}$ as the 
function of the lightest neutrino mass eigenvalue $m_{\rm ref}$. It shows that 
the ratio becomes relatively large (small) if the neutrino mass spectrum is 
NH (IH). With being large value of $m_{\rm ref}$, namely QD mass spectrum 
limit, $\hat{m}$ takes similar values for normal and inverse ordering cases. 

In the tri-bimaximal limit, ${\cal B}(\mu\to e\gamma)$ and 
${\cal B}(\tau \to e\gamma)$ in~\eqref{lfvtbl} are much suppressed than 
${\cal B}(\tau\to \mu\gamma)$ since they do not involve $m_{123}$. This 
implies that these processes are sensitive to deviations from the 
tri-bimaximal mixing. As argued in Section~\ref{sec:Neu}, the deviation 
parameters in~\eqref{mnstb} are sufficiently small, so that we 
can use them as expansion parameters in LFV branching fractions. Up to 
${\cal O}(\epsilon_w^2)$, one can obtain the following expressions: 
  \begin{eqnarray}\notag
  |b_{12}|^2&\simeq &\tilde m_{12}^2+\sqrt{2}\tilde m_{12}^2\epsilon_x
-2\tilde m_{12}^2\epsilon_y+{\tilde m_{12}\over\sqrt{2}}(\tilde m_{12}+m_{123})
\epsilon_z\cos\delta -{7\over 2}\tilde m_{12}^2\epsilon_x^2
-2\sqrt{2}\tilde m_{12}^2\epsilon_x\epsilon_y\\\label{c122}
&&
-{1\over 8}(7\tilde m_{12}^2-2\tilde m_{12}m_{123}-m_{123}^2)\epsilon_z^2
-{\tilde m_{12}\over 2}(7\tilde m_{12}-m_{123})\epsilon_x\epsilon_z\cos\delta,\\
    \notag
  |b_{23}|^2&\simeq &{1\over 16}(\tilde m_{12}-m_{123})^2
-{\tilde m_{12}\over\sqrt{2}}(\tilde m_{12}-m_{123})\epsilon_x
+{\tilde m_{12}\over 4}(7\tilde m_{12}+m_{123})\epsilon_x^2
-{1\over 4}(\tilde m_{12}-m_{123})^2\epsilon_y^2\\\label{c232} 
&&
+{1\over 8}(\tilde m_{12}^2-m_{123}^2+16\tilde m_{12}^2\sin^2\delta)\epsilon_z^2
-\sqrt{2}\tilde m_{12}(\tilde m_{12}-m_{123})\epsilon_y\epsilon_z\cos\delta, 
\end{eqnarray}
%
where $\tilde m_{12}\equiv m_{12}/3$ and 
$|b_{13}|^2(\epsilon_y, \epsilon_z\cos\delta)
=|b_{12}|^2(-\epsilon_y, -\epsilon_z\cos\delta)$. 

{}From the fact that $m_{123}$ is much larger than $m_{12}$, the typical 
correlation between neutrino parameters and the branching fractions 
can be understood. In~\eqref{c122}, the terms which remain in the 
tri-bimaximal limit are proportional to $m_{12}^2$, and $m_{123}$ always 
appears with involving the deviation parameters, especially $\epsilon_z$.  
Thus $|b_{12(13)}|^2$, namely ${\cal B}(\mu(\tau)\to e\gamma)$, is 
sensitive to the parameters, while ${\cal B}(\tau\to \mu\gamma)$ does not 
have so large dependence on them since $m_{123}^2$ is the dominant term in 
$|b_{23}|^2$. Moreover, the leading contributions in $|b_{12}|^2$ and 
$|b_{23}|^2$ can be expressed as 
\begin{equation}
|b_{12}|^2\;\simeq\; 
{m_{123}^2\over 9}
\left(
\hat m^2+{3\over \sqrt{2}}\hat m \epsilon_z \cos \delta+{9\over 8}\epsilon_z^2
\right)+\cdots,
\qquad  
  |b_{23}|^2\;\simeq\;
{m_{123}^2\over 16}+\cdots, 
\label{ld}
\end{equation}
since the ratio $\hat{m}$ is also the small quantity as well as 
deviation parameters. One can find that $\epsilon_z$ plays  an 
important role to determine ${\cal B}(\mu(\tau)\to e\gamma)$ and that 
$\epsilon_x$ and $\epsilon_y$ are less effective  because they only 
appear as sub-leading corrections.  The $\epsilon_z$ dependence 
is controlled by the neutrino mass spectrum and the Dirac phase 
through $\hat{m}$ and $\cos\delta$.  For example, $|b_{12}|^2$ is minimized at 
\begin{eqnarray}
 \epsilon_z&=&\theta_{13}\;\simeq\;-{2\sqrt{2}\over 3}\hat m\cos\delta 
 \;\simeq\;-\hat m\cos\delta. 
\label{ez-min}
\end{eqnarray}
Note that the value of $\epsilon_z$ in~\eqref{ez-min} has significant 
distinction between NH and IH by the magnitude 
of $\hat m$.  We can check these properties  with numerical analysis. 

\paragraph{Numerical searches}

We discussed the neutrino parameter dependence of the LFV prediction using 
the approximate formula in~\eqref{brap}. To make our study more complete, 
we proceed to numerical examination of the LFV prediction.  Here we take 
SPS1$_a$~\cite{sps1a} for SUSY particle mass spectrum; SUSY breaking 
parameters are fixed as $(m_0,M_{1/2},A_0)=(100,250,-100)$~GeV at the GUT scale 
and $\tan\beta=10$. SUSY parameter dependence is mostly flavor blind, and 
has been greatly studied~\cite{lfv}. If one takes other types of SUSY mass 
spectrum, following results do not alter as long as the universality of the 
SUSY breaking is assumed. 

Given set of SUSY parameters, we numerically estimate RG evolutions 
between GUT and electroweak scale taking heavy right-handed neutrinos 
into account. Above the right-handed neutrino mass scale, which is taken 
as $M_U=10^{14}$~GeV in the analysis, the right-handed neutrinos are decoupled 
with the theory.  Two-loop RG equations for gauge and Yukawa couplings, and 
one-loop ones for the soft SUSY breaking parameters are numerically solved. 
LFV fractions are estimated with one-loop diagrams in the SUSY particle mass 
eigenbasis rather than the mass-insertion approximation. 

\begin{figure}[t]
\begin{center}
  \includegraphics[width=15.cm,clip]{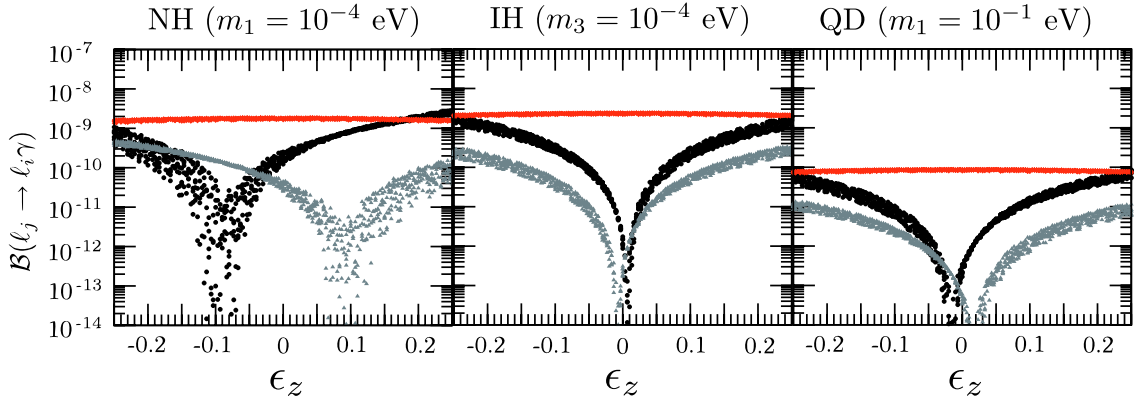}
\end{center}
\caption{Predictions of LFV branching fractions as the functions of 
$\epsilon_z$ are shown. Black, red and gray plots correspond to 
${\cal B}(\mu\to e\gamma)$, ${\cal B}(\tau\to \mu\gamma)$ and 
${\cal B}(\tau\to e\gamma)$, respectively. Each figure corresponds to 
different neutrino mass spectrum: from left to right figures, NH 
($m_1=10^{-4}$~eV), IH ($m_3=10^{-4}$~eV), and QD ($m_1=10^{-1}$~eV) mass 
spectra are taken, respectively. The mass squared differences are fixed to 
central values in~\eqref{msqdiff} and the Dirac phase is taken as $\delta=0$. 
$\epsilon_x$ and $\epsilon_y$ are scanned with $3\sigma$ ranges 
in~\eqref{devrange}. 
  \bigskip}
\label{ez}
\end{figure}

As stressed in the analytical discussion, among the three deviation 
parameters in~\eqref{devrange}, $\epsilon_z$ is crucial for the prediction 
of $\mu(\tau)\to e\gamma$ branching fractions. To see its dependence, we 
plot the LFV predictions as the functions of $\epsilon_z$ in Fig.~\ref{ez}. 
The prediction of ${\cal B}(\tau\to \mu\gamma)$ is insensitive to size 
of $\epsilon_z$, by contrast that of ${\cal B}(\mu(\tau)\to e\gamma)$ strongly  
depends on. The branching fraction ${\cal B}(\mu\to e\gamma)$ 
is minimized around $\epsilon_z\simeq -0.1 (+0.01)$ for the case with NH 
(IH and QD) mass spectrum, as shown in \eqref{ez-min}. 
These results are consistent with the previous argument using the analytical 
expressions. Note that ${\cal B}(\mu(\tau)\to e\gamma)$ is highly suppressed 
than ${\cal B}(\tau \to \mu\gamma)$ around the minima. 
This is an important point to extract implications for neutrino parameters from 
future LFV searches, and we discuss the issue in the next section. 

\paragraph{Effects of heavy Majorana mass hierarchies}

In general, heavy Majorana masses have non-degeneracy and it affects  
LFV prediction. In order to incorporate the non-degeneracy of $M_R$, it 
is convenient to focus on $B_{ij}$ in~\eqref{fl} rather than $b_{ij}$ 
in~\eqref{cij}.  $B_{ij}$ can be generally rewritten as follows~\cite{rdep}: 
\begin{eqnarray}
B_{ij}
&=&
\sum_{k=1}^3 
(U^*  {\cal D}_{\sqrt{m}}R{\cal D}_{\sqrt{M_R}})_{ik}
({\cal D}_{\sqrt{M_R}}R^\dag{\cal D}_{\sqrt{m}}U^T)_{kj}
\ln{M_G\over M_{Rk}}, 
\label{yygen}
\end{eqnarray}
where ${\cal D}_{\sqrt m}=diag(\sqrt{m_1},\sqrt{m_2},\sqrt{m_3})$, 
${\cal D}_{\sqrt {M_R}}=diag(\sqrt{M_{R1}},\sqrt{M_{R2}},\sqrt{M_{R3}})$ 
and a complex matrix $R$ satisfies $RR^T={\bf 1}$.  The additional 
mixing matrix $R$ appears in $Y_\nu Y_\nu^\dag$ because the right-handed 
neutrino mixing is not unphysical. As a result, LFV prediction generally 
depends on the mixing structure of $R$, as minutely studied in~\cite{rdep}. 

If the right-handed mixing is not important for low-energy neutrino 
parameters, namely $R\simeq {\bf 1}$, $B_{ij}$ is simplified. Especially 
$R\to {\bf 1}$ leads to 
\begin{eqnarray}
  B_{ij}&\simeq&\sum_{k=1}^3
{\cal M}_{k}
(U^*)_{ik}
(U^T)_{kj}
, \qquad 
{\cal M}_i\;\equiv \;
m_iM_{Ri}\ln{M_G\over M_{Ri}}.
\end{eqnarray}
It is obvious that $B_{ij}$ has a similar form to $b_{ij}$, where 
low-energy neutrino mass $m_i$ in~\eqref{cij} is replaced by a combination 
of light and heavy neutrino masses. Hence, explicit expression 
of $B_{ij}$ is easily obtained by~\eqref{cijfull}. In this case, LFV 
prediction depends on both neutrino masses through 
${\cal M}_{12}={\cal M}_2-{\cal M}_1$ and 
${\cal M}_{123}=2{\cal M}_3-{\cal M}_2-{\cal M}_1$. In particular, the new 
mass ratio $\hat {\cal M}={\cal M}_{12}/{\cal M}_{123}$ is essential to 
determine the neutrino parameter dependence. With typical mass hierarchies 
of $M_R$, LFV prediction has examined in Ref.~\cite{MReff}. 
Corrections to $R={\bf 1}$ also affect LFV prediction when right-handed 
neutrinos have large non-degeneracy. For example, if $R$ matrix 
in~\eqref{yygen} contains a small mixing angle 
$\kappa_{ij}$,\footnote{Here $\kappa_{ij}$ is assumed to 
be real. Complex phases of $R$ can bring further modification into LFV 
prediction~\cite{rdep}.} then $\mu\to e\gamma$ prediction is modified with 
including the counterpart of~\eqref{ld}; up to ${\cal O}(\kappa_{ij})$ 
contribution except for ${\cal O}(\kappa_{ij}\cdot\epsilon_{x,y,z})$ 
terms, the prediction is explicitly written as follows: 
\begin{eqnarray}
  |B_{12}|^2\;\simeq \; {{\cal M}_{123}^2\over 9}
\bigg[\hat{\cal M}^2+{3\over \sqrt{2}}\hat{\cal M} \epsilon_z\cos\delta
+{9\over 8}\epsilon_z^2
+{\hat{\cal M}}
{\sqrt{2}\Delta_{12}+\sqrt{6}\Delta_{23}+2\sqrt{3}\Delta_{13}
\over {\cal M}_{123}}
+\cdots
\bigg],
\label{B12sq}
\end{eqnarray}
where we use the notation 
\begin{eqnarray}
  R\;=\;
  \begin{pmatrix}
    1&\kappa_{12}&\kappa_{13}\\
    -\kappa_{12}&1&\kappa_{23}\\
    -\kappa_{13}&-\kappa_{23}&1
  \end{pmatrix},\quad 
\Delta_{ij}\;\equiv \;\kappa_{ij}\sqrt{m_im_j}\left(
M_{Rj}\log{M_G\over M_{Rj}}-M_{Ri}\log{M_G\over M_{Ri}}\right). 
\end{eqnarray}
The contribution due to $R\neq {\bf 1}$ strongly depends on heavy Majorana mass 
hierarchy. 
One can similarly see that the leading contribution from  $\kappa_{ij}$ 
appears in $\Delta_{ij}$ for the $\tau\to e\gamma$ and $\tau\to \mu\gamma$ 
predictions. 

The parameter dependence of the branching fractions is modified from 
the degenerate heavy Majorana case by mainly given difference between $\hat m$ 
and $\hat {\cal M}$. Nevertheless, a particular Majorana mass spectrum is taken 
such as $M_{R3}$ is dominantly heavy, then similar discussion to the 
degenerate case is possible as long as effects from $R$ in~\eqref{yygen} is 
sufficiently small.\footnote{In a class of models where all the leptonic 
flavor violation originates in the charged lepton sector correspond to 
$R={\bf 1}$ as in Ref.~\cite{rdep}. This is equivalent to the case that 
neutrino Yukawa and heavy Majorana mass matrices can be taken as 
simultaneously diagonal. It is notified that such the situation is 
approximately realized in some $E_6$ and $SO(10)$ grand unified models, 
called lopsided mass structure~\cite{lops}.}
Further study on effects of $R$ and Majorana masses from our viewpoint is 
also important and future task. 

\section{Probing neutrino parameters with LFV searches}
\label{sec:clarif}

Finally, we investigate possible implications for neutrino parameters from 
future LFV searches with the analysis obtained in the previous section. 
In the following we concentrate on a limited scenario where $R={\bf 1}$ 
is assumed, and show how future LFV searches give constraints for neutrino 
mass spectrum and $\theta_{13}=\epsilon_z$. 

Experimental discovery of lepton rare decay processes 
$\ell_j\to\ell_i+\gamma$ is one of smoking gun signals of physics beyond 
the SM; thus several experiments have been developed to detect LFV processes. 
The present experimental upper bounds are given 
at 90\% C.L.~\cite{muex,tauex}: 
\begin{eqnarray}\notag
  {\cal B}(\mu\to e\gamma)&\leq &1.2\times 10^{-11},\quad
  {\cal B}(\tau\to \mu\gamma)\;\leq \;4.5\times 10^{-8},\quad 
  {\cal B}(\tau\to e\gamma)\;\leq \;1.2\times 10^{-7}. 
\end{eqnarray}
These bounds are to be modified in near future searches. MEG experiment 
searches $\mu\to e\gamma$; the bound is expected to reach 
${\cal B}(\mu\to e\gamma)\leq {\cal O}(10^{-13}\sim 10^{-14})$~\cite{meg}. 
Future $B$-factories would also greatly reduce the $\tau$ decay upper 
bounds~\cite{superb}. In our analysis, we conservatively adopt 
\begin{eqnarray}
  {\cal B}(\mu\to e\gamma)&\lesssim &10^{-13},\quad
  {\cal B}(\tau\to \mu\gamma)\;\lesssim \;10^{-9},\quad 
  {\cal B}(\tau\to e\gamma)\;\lesssim \;10^{-9},  
\label{future}
\end{eqnarray}
as upcoming upper bounds of LFV fractions. 
Since the bound for ${\cal B}(\mu\to e\gamma)$ is most severe, 
${\cal B}(\mu\to e\gamma)/{\cal B}(\tau\to\mu(e)\gamma)$ must be 
sufficiently suppressed as $10^{-2}\sim 10^{-5(6)}$ in order to 
observe both $\mu$ and $\tau$ decay processes. 
Prediction of ${\cal B}(\tau\to e\gamma)$ is always suppressed than 
${\cal B}(\tau\to \mu\gamma)$ in~Fig.~\ref{ez}, and we focus on 
$\tau\to \mu\gamma$ between the tau decay processes in the following.

\begin{figure}[t]
\begin{center}
  \includegraphics[width=14.cm,clip]{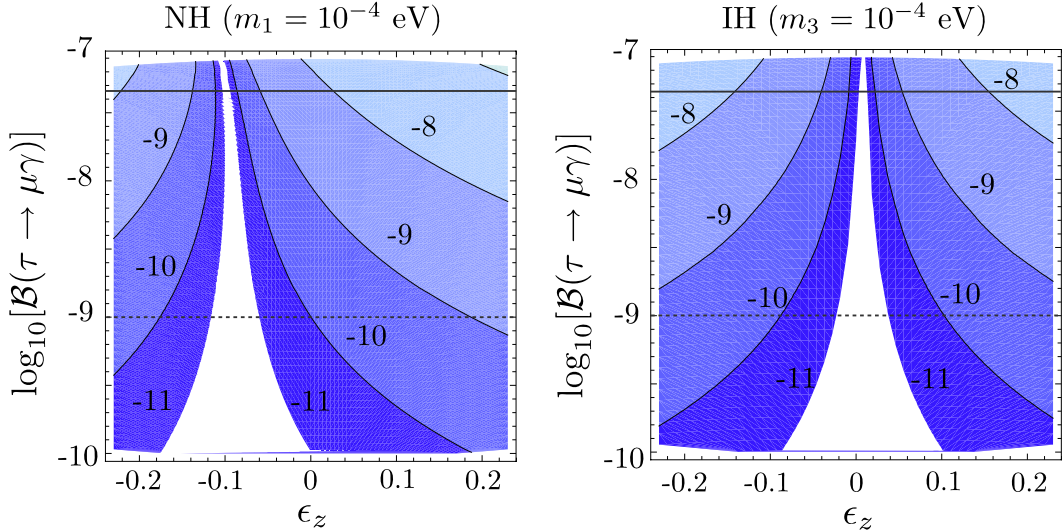} 
\end{center}
\caption{Contour plots of ${\cal B}(\mu\to e\gamma)$ as the functions 
of $\epsilon_z$ and ${\cal B}(\tau\to \mu\gamma)$ for the cases with 
NH and IH. The other deviation parameters are set to zero, 
and the neutrino masses are fixed to their central values 
in~\eqref{msqdiff}. The Dirac phase is taken as $\delta=0$. 
Shaded regions have already been excluded by the current experimental bound for 
$\mu\to e\gamma$, and the current (expected) bounds for $\tau\to \mu\gamma$ 
is shown by the solid (dotted) lines. 
  \bigskip}
\label{ez2}
\end{figure}

The LFV fractions depend on $M_R$ and SUSY breaking parameters in a flavor 
blind way, and on the neutrino parameters in a flavor dependent manner. As 
seen in the previous section, predictions of $\mu\to e\gamma$ and 
$\tau\to e\gamma$ highly depend on $\epsilon_z$, but $\tau\to \mu \gamma$ is 
nearly independent of it. Thus ${\cal B}(\tau\to \mu \gamma)$ is mostly 
determined by $M_R$ and SUSY breaking parameters; namely, the universal 
dependence in LFV fractions can be read from ${\cal B}(\tau\to \mu \gamma)$. 
Hence we can express ${\cal B}(\mu\to e\gamma)$ using $\epsilon_z$ and 
${\cal B}(\tau\to\mu\gamma)$.
Fig.~\ref{ez2} shows contour plots of ${\cal B}(\mu\to e\gamma)$ as 
the function of $\epsilon_z$ and ${\cal B}(\tau\to \mu\gamma)$ for the 
cases with NH and IH. In the figure, shaded regions have already been 
excluded by the current experimental bound for $\mu\to e\gamma$, and 
the current and expected bounds for $\tau\to \mu\gamma$ are shown by 
the solid and dotted lines, respectively. 

{}From the figure, one can realize that future LFV searches give us 
implications for neutrino parameters. For instance, if near future experiments 
discover both ${\cal B}(\mu\to e\gamma)$ and ${\cal B}(\tau\to \mu\gamma)$, 
then $\epsilon_z$ and neutrino mass spectrum are strongly constrained. In 
the scenario, on the one hand for NH case $|\theta_{13}|$ is close 
to 0.1, on the other hands for IH and QD cases such a large value of 
$\theta_{13}$ is not allowed. It is interesting that the above value of 
reactor angle for NH is in accord with the best-fit value reported by 
recent data analyses. Hence, the LFV discovery excludes IH and QD mass spectra 
when the large  $\theta_{13}$ is established in experiments like 
T2K~\cite{t2k} and Double Chooz~\cite{dch}. Though in the analysis 
the Dirac phase is taken as zero, the result is preserved if non-zero value 
of $\delta$ is incorporated. This is because IH and QD spectra 
are still inconsistent with the LFV discovery and the large $\theta_{13}$ 
value, as easily realized from \eqref{ez-min}. 
\begin{figure}[t]
\begin{center}
  \includegraphics[width=10.cm,clip]{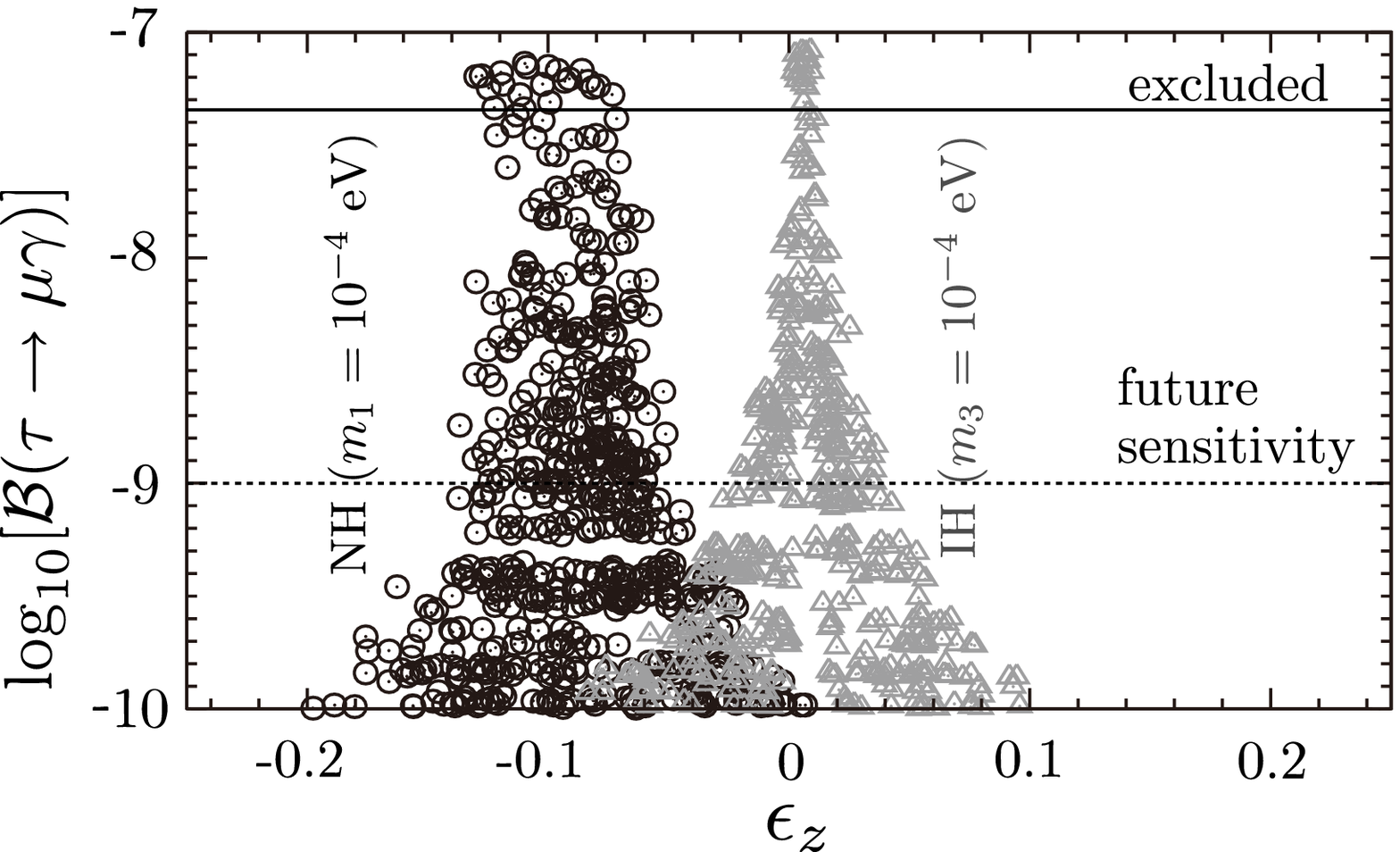} 
\end{center}
\caption{Prediction of ${\cal B}(\tau\to \mu\gamma)$ as the functions 
of $\epsilon_z$ for the cases with NH (black circles) and IH 
(gray triangles). All the plotted points satisfy 
${\cal B}(\mu\to e\gamma)\leq 1.2\times 10^{-11}$. 
The deviation parameters $\epsilon_x$ and $\epsilon_y$ are scanned with 
3$\sigma$ ranges~\eqref{devrange}, and the Dirac phase is taken as $\delta=0$. 
The current (expected) bounds for $\tau\to \mu\gamma$ is shown by the solid 
(dotted) lines. 
  \bigskip}
\label{scan3}
\end{figure}

As another scenario, when future LFV searches discover 
only ${\cal B}(\tau\to \mu\gamma)$, the above discussion is 
still valid. Fig.~\ref{scan3} shows the prediction of 
${\cal B}(\tau\to \mu\gamma)$ as the functions of $\epsilon_z$ for the cases 
with NH and IH. All the plotted points satisfy 
${\cal B}(\mu\to e\gamma)\leq 1.2\times 10^{-11}$. One can see that NH and IH 
require different values of $\epsilon_z$ to predict 
${\cal B}(\tau\to \mu\gamma)$ in future discovery range.
However, constraints on $\epsilon_z$ and neutrino mass spectrum 
are weakened, if ${\cal B}(\tau\to \mu\gamma)$ is sufficiently 
suppressed than the future experimental limit. 

\section{Conclusion}
\label{sec:conclusion}

In this Letter, we have examined relation between neutrino parameters 
and LFV predictions, in light of the tri-bimaximal mixing and the recent 
precision data. By using a particular parametrization for the lepton mixing 
matrix, which is useful to study difference between the MNS and tri-bimaximal 
mixing matrices, we have explicitly showed that the flavor dependence in LFV 
predictions is controlled by deviation parameters and neutrino mass 
differences. 

In the setup with universal heavy Majorana masses and soft SUSY breaking 
parameters, we have found that $\epsilon_z$ and the neutrino mass 
spectrum are important for predictions of ${\cal B}(\mu\to e\gamma)$ and 
${\cal B}(\tau\to e\gamma)$, while $\epsilon_x$ and $\epsilon_y$ are less 
effective to determine the LFV predictions. The branching fraction 
${\cal B}(\tau\to \mu\gamma)$ is also shown to be insensitive to 
neutrino parameters. In addition, we have discussed the effects from heavy 
Majorana mass structure, namely non-degeneracy in right-handed neutrinos and 
small mixing angles in $R$ matrix. 

We have examined and extracted the possible implications for neutrino 
parameters from upcoming LFV searches. Future discovery of LFV process can 
give strong constraints on $\theta_{13}$ with respect to the type 
of neutrino mass hierarchy as long as effects from $R$ in~\eqref{yygen} is 
sufficiently small. In particular, $\tau\to\mu\gamma$ discovery 
excludes IH and QD neutrino mass spectra if $|\theta_{13}|\simeq 0.1$ 
would be established. 

In general, inclusive studies of the precision neutrino parameters and LFV 
could give us implications for unrevealed issues in the lepton flavor 
structure, such as the neutrino tri-bimaximality. Further investigations of 
LFV prediction focusing on effects from the phases and the Majorana mass 
structure are our future works. 

\bigskip
\subsection*{Acknowledgments}
The authors would like to thank M.~Tanimoto for helpful discussions, 
and K.~Inoue, K.~Harada, H.~Kubo, N.~Yamatsu and S.~Kaneko for fruitful 
comments. 
This work was supported by a Grant-in-Aid 
for Scientific Research on Priority Areas (\#441) ``Progress in 
elementary particle physics of the 21st century through discoveries of 
Higgs boson and supersymmetry''(No.~16081209) from the Ministry of 
Education, Culture, Sports, Science and Technology of Japan.



\end{document}